\documentclass{jfm}
\usepackage{graphicx}
\usepackage{epstopdf, epsfig}
\usepackage{amsfonts,amssymb,amsmath,mathbbol} 
\usepackage[breaklinks,hidelinks]{hyperref}

\usepackage{xcolor} 

\usepackage{appendix}

\usepackage{color} 

\shorttitle{Cracking of submerged beds} 
\shortauthor{Bhadra et al.} 

\title{Cracking of submerged beds}

\author
 {
 Satyanu Bhadra,\aff{1}
 Anit Sane,\aff{1}
 Akash Ghosh,\aff{1}
 Shankar Ghosh,\aff{1}
 \and
 Kirti Chandra Sahu\aff{2}\corresp{\email{ksahu@che.iith.ac.in, sghosh@tifr.res.in}}
 }

\affiliation
{
\aff{1}
Department of Condensed Matter Physics and Materials Science, Tata Institute of Fundamental Research, Mumbai 400005, India
\aff{2}
Department of Chemical Engineering, Indian Institute of Technology Hyderabad, Sangareddy 502 284, Telangana, India
}

\begin{document}

\maketitle

\begin{abstract}
We investigate the phenomena of crater formation and gas release caused by projectile impact on underwater beds, which occurs in many natural, geophysical, and industrial applications. The bed in our experiment is constructed of hydrophobic particles, which trap a substantial amount of air in its pores. In contrast to dry beds, the air-water interface in a submerged bed generates a granular skin that provides rigidity to the medium by producing skin over the bulk. The projectile's energy is used to reorganise the grains, which causes the skin to crack, allowing the trapped air to escape. The morphology of the craters as a function of impact energy in submerged beds exhibits different scaling laws than what is known for dry beds. This phenomenon is attributed to the contact line motion on the hydrophobic fractal-like surface of submerged grains. The volume of the gas released is a function of multiple factors, chiefly the velocity of the projectile, depth of the bed and depth of the water column. 
\end{abstract}



\section{Introduction} \label{sec:intro}

The release of gas bubbles from a submerged granular bed is important in many natural, geophysical, and industrial applications \citep{gostiaux2002dynamics}. For example, the release of methane bubbles held in the ocean and lake beds due to the decomposition of organic matter can have serious environmental consequences \citep{meier2011bubbles}. Similarly, in the context of carbon capture and storage in seabeds, a significant concern arises regarding the potential escape of CO$_2$ bubbles into the environment due to the rupture of storage sites \citep{sellami2015dynamics}. Therefore, it is essential to understand the different means that can lead to such an adverse situation, one of which is the impact of an object on a submerged granular bed. Apart from the wide range of practical applications, this subject involves complex physics, such as the formation of craters when an object collides with a granular bed. Specifically, it is accompanied by a number of complicated events, including shock wave propagation, compaction, granular material ejection, contact line dynamics, and rearrangement. All these events can substantially impact the final shape and size of the formed craters \citep{melosh1989impact}, which, in turn, facilitates trapped air inside the bed to escape in the form of air bubbles.

In dry conditions, the Froude Number, $Fr= v^2/2gD_p$, can be used to classify the craters \citep{walsh2003,holsapple1993scaling}. Here, $v$ denotes the impact velocity of the object, $g$ is the acceleration due to gravity, and $D_p$ represents the object's diameter. The impact of an object on dry granular media has been investigated over a wide range of $Fr\in (10,10^7)$. The craters are usually characterized by the ratio between the diameter $(D_C)$ and the crater depth $(h)$. Simple parabolic-shaped craters with well-defined rims and sharp crests are formed when $Fr<10^4$. For these craters, $D_C/h$ is $\sim 5$. In this parameter regime, it has been established that $D_C \sim E_k^{1/3}$, if the strength of the impacted compressed surface dominates the crater formation process. For impacts in loose granular surfaces, the impact energy ($E_k$) is spent on ejecting and depositing the substrate material \citep{villalobos2022geometrical}. For such cases, $D_C \sim E_k^{1/4}$ and $h \sim D_P^{-5/6} E_k^{1/3}$ \citep{PhysRevLett.90.194301}. Similar behaviour of $h \sim E_k^{1/3}$ has been observed for impacts on surfaces made of dust aggregates \citep{katsuragi2017physics}, glass beads \citep{de_vet2007shape,katsuragi2007unified}, and in the collapse of granular columns \citep{de_vet2012collapse}. On the other hand, for $Fr>10^4$, complex craters develop, which are characterized by a high aspect ratio, $D_C/h \sim 10-20$ \citep{kruger2018deriving}. The complex craters also feature flat interior floors, secondary internal rings, and central peak-like morphology.

In wet conditions, the impact of an object in a submerged granular bed introduces additional complexity, characterized by the formation of capillary bridges and contact line dynamics, imposing extra rigidity on the system \citep{capillarybridge1,cap3soulie2006influence,cap4herminghaus2005dynamics}. Experiments with solid projectiles impacting a wet granular surface reveal three types of craters: (i) simple craters (similar to that seen in dry sand media) for low levels of water saturation, (ii) transitional craters (with a steeper wall than simple craters) for intermediate levels of water saturation, and (iii) cylindrical craters for high levels of water saturation. For simple and transitional craters, $h \sim E_k^{1/5}$ \citep{partiallywetsandPhysRevE.88.022203}. This scaling breaks down in the case of cylindrical craters. Thus, the impact of a projectile with a granular bed initiates a compaction process. In dry granular systems, this compaction is resisted by frictional forces, while in wet or submerged granular systems, contact line forces come into play to counteract the compaction. In a submerged granular system, the compaction process causes the rearrangement of particles, and in some cases, leads to the release of the contained gas.

This present investigation addresses the release of gas from underwater beds by examining crater formation in hydrophobic sand beds submerged with $Fr$ in the range of 1 to 100. In these beds, low-density air is trapped below high-density water. This unstable configuration is made possible by pinning the three-phase contact line comprising trapped air, water, and solid sand grains coated with a hydrophobic material \citep{liquidmarbleC1SM05066D,krm17aussillous2006properties,krm18abkarian2013gravity,krm19subramaniam2006mechanics,krm20pakpour2012construct,karmakargsPhysRevE.95.042903}. As the projectile impacts the sand bed, this unstable configuration is disturbed, releasing trapped air and crater formation. The present study reveals that the volume $(V)$ of air released upon impact is proportional to the kinetic energy ($E_k$) of the impactor and is influenced by the height ($L_W$) of the liquid column above the sand bed. For low-energy impacts, the crater depth ($h$) scales as $E_k^{1/2}$, and its diameter ($D_C$) scales as $E_k^{1/4}$. Furthermore, our study suggests that the new scaling laws obtained at a low $E_k$ are associated with the depinning of the three-phase contact line, which is attributable to the fractal nature of the hydrophobic surface. A model based on interfacial tension associated with an increase in the surface area of the submerged sand bed due to the impact explains the experimental observations.

\begin{figure}
\centering
\includegraphics[width=0.9\linewidth]{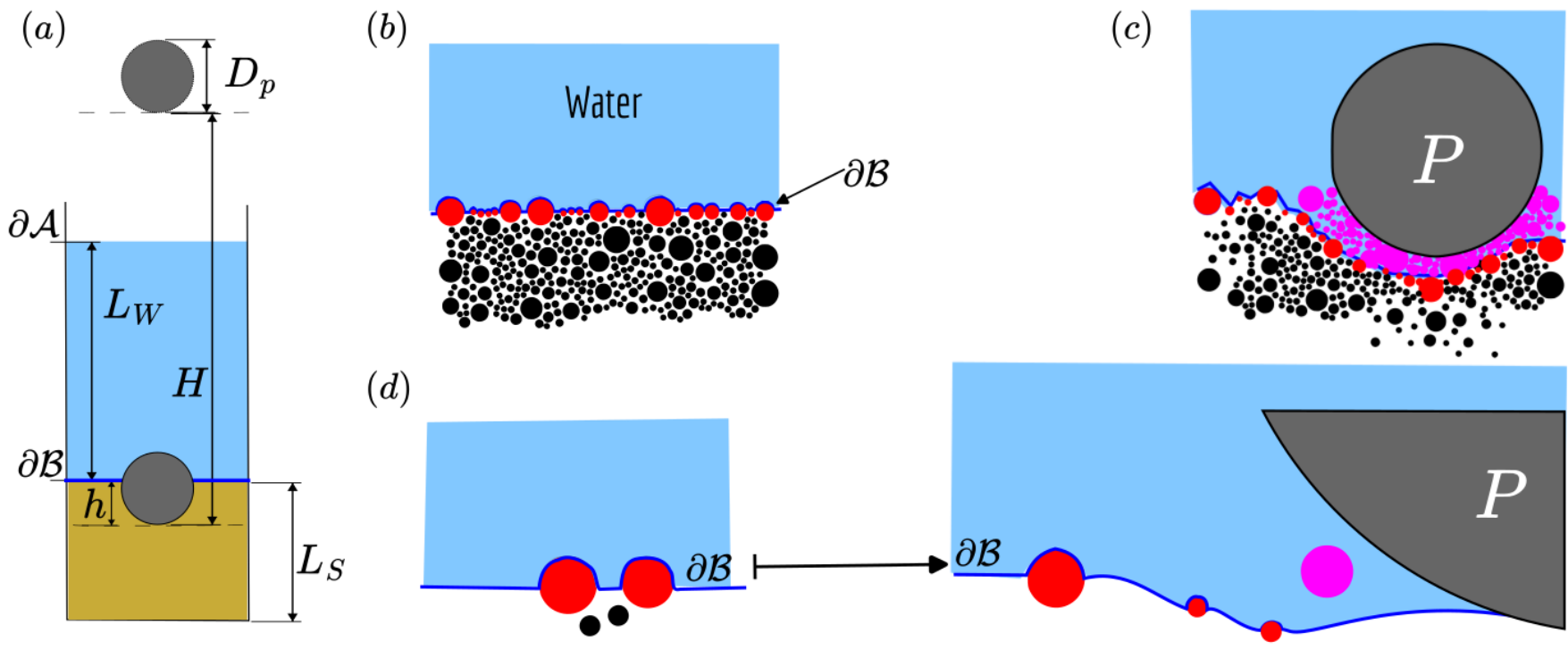}
\caption{(a) The geometrical representation of the experiment. Here, $L_S$, $L_W$, $H$ and $h$ are the sand bed's height, the water column's height, the vertical distance travelled by the projectile and the crater's depth, respectively. $\partial \mathcal{A}$ and $\partial \mathcal{B}$ represent the top air-water interface and the air-water-sand interface on the granular bed, respectively. (b,c,d) Schematic representation of the interface depinning during the collision. (b) The boundary between the solid and liquid phases ($\partial B$; represented by a blue line) separates trapped air from the liquid pool. (c) Upon the impact of a projectile ($P$) on this interface, particle arrangement is disrupted, leading to the reconfiguration of $\partial B$. The magenta-coloured particles are now submerged in water. (d) An illustrative example of particle rearrangement causing interface reconfiguration. It shows that following the collision, the interface shifts towards the two black particles, resulting in the encapsulation of the rightmost particle by water. In panels (b,c,d), grains on the interface are highlighted in red, while those in the dry interior are marked in black.}
\label{fig:statement}
\end{figure}

Figure \ref{fig:statement}(a) illustrates the experimental geometrical parameters. A hydrophobic sand bed with a height $L_S$ is submerged within a water pool. As discussed earlier, the impact of the projectile induces a dynamic compaction of the granular bed. The mechanical integrity of the bed is compromised when the height of the sand column, $L_s$, is comparable to the size of the projectile, $D_p$. Thus, we have performed experiments for $L_s \gg D_p$. The height of the water pool above the sand bed is denoted as $L_W$. The experiments involve dropping projectiles ($P$) with varying diameters ($D_p$) from different heights ($H$) into the granular bed. As the projectile descends, it interacts with two interfaces: $\partial \mathcal{A}$ representing the air-water interface and $\partial \mathcal{B}$ representing the three-phase (air-water-sand) contact line. Upon colliding with the granular bed, we scrutinize the changes in the volume ($V$) of air released from the bed and the shapes of the craters in relation to the impact kinetic energy $(E_k)$.

Before discussing the details of the experiments, we highlight the core issue that underscores the connection between the pinning of the three-phase (air-water-sand) contact line and crater formation. Figure \ref{fig:statement}(b) schematically illustrates the hydrophobic sand trapping air within its pore spaces. This is accomplished by forming a protective skin ($\partial \mathcal{B}$ as shown in Figure \ref{fig:statement}(b)) that stabilizes the trapped air bubble against buoyancy forces. The contact lines at the air-water-sand interface, constituting the `skin', are immobilized (pinned) by defects on the sand grains. The skin has a rigid surface with an effective elastic constant 
\citep{protiere2023particle,karmakargsPhysRevE.95.042903}. Upon impact by the projectile, the sand grains undergo displacement from their original positions. This displacement is resisted by capillary forces stemming from the arrangement of the three-phase contact lines. If these restoring forces are surpassed, the interface depins, and the sand grains experience displacement. Consequently, a new interface is established, allowing water to penetrate the granular assembly during this process. Figure \ref{fig:statement}(c) provides a schematic representation of this reconfiguration. At the microscopic level, a depiction of this reconfiguration involving a few sand grains is presented in Figure \ref{fig:statement}(d), where the left image serves as the reference. 

\begin{figure}
\centering
\includegraphics[width=0.9\linewidth]{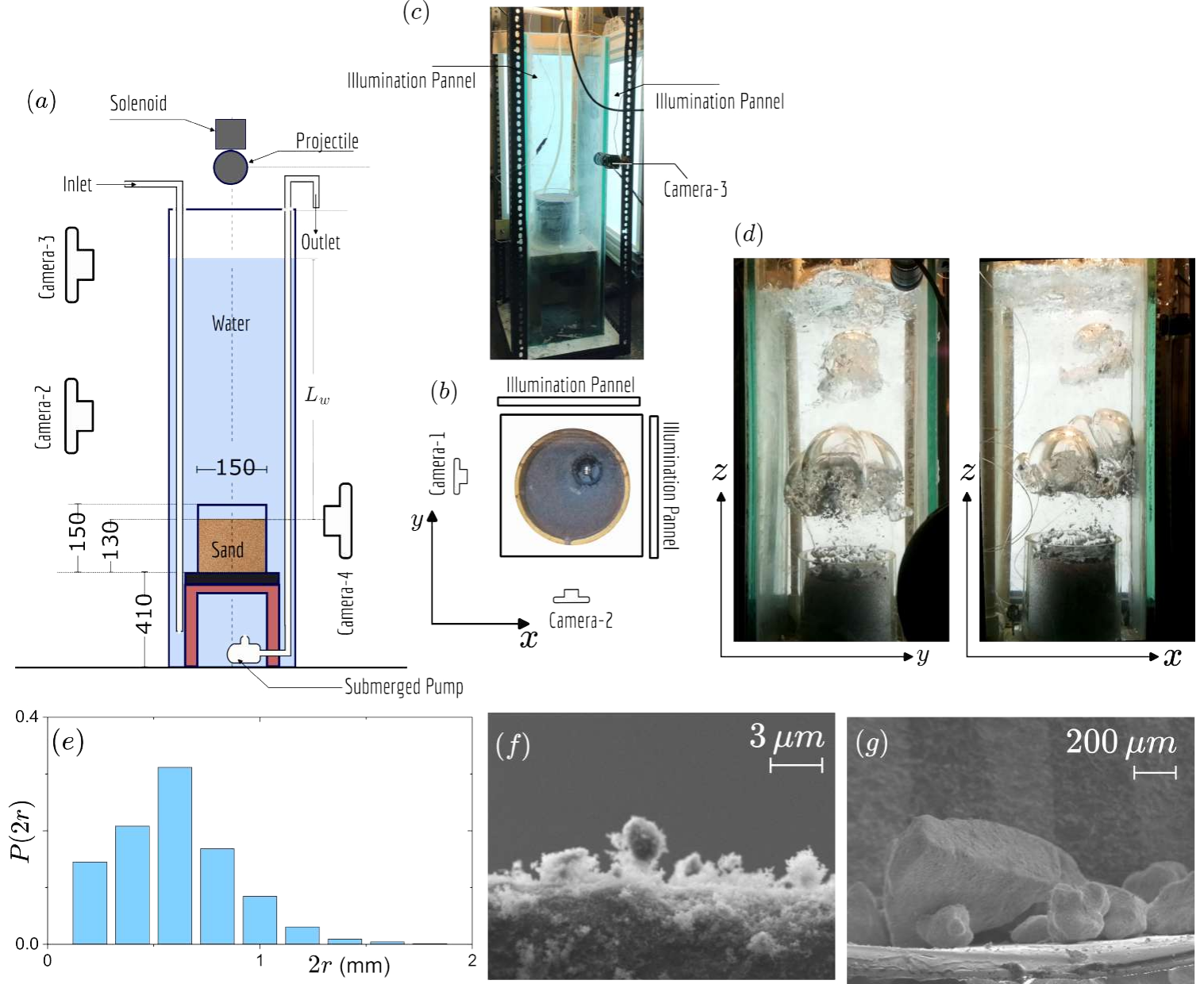}
\caption{Schematic representation of the experimental setup: (a) side view and (b) the top view $(xy)$. The relevant dimensions are in $\textrm{mm}$. As seen in the figure, we use spherical projectiles (stainless steel, density 7600 kg/m$^3$) of various diameters. (c) The actual experimental set-up. (d) Representative images of the escaping bubbles after impact in the $yz$ and $xz$ planes. (e) Size distribution of the sand grains. (f) A magnified image of the grains showing the hydrophobic coating and (g) the polyhedral nature of the individual grains captured using scanning electron microscopy (SEM).}
\label{fig:Figure1_schematic}
\end{figure}

\section{Experimental procedure} \label{sec:expt}

Figure \ref{fig:Figure1_schematic}(a) illustrates the experimental setup employed in the present study. Hydrophobic sand is poured into a cylindrical container with dimensions of 150 mm in diameter and 150 mm in height, leaving a 20 mm gap, thus, $L_S=130$ mm. The importance of the gap of 20 mm will be discussed later while discussing the imaging techniques. The sand is poured freely into the container, followed by tapping and compactifying the grains, akin to the coffee-tamping process. This method ensures the absence of loose grain clusters within the sand bed and approximately traps 38\% air by volume in the pores. The entire assembly is then lowered into a rectangular glass container on a raised platform using a pulley system, as depicted in the side view of the experimental setup (Figure \ref{fig:Figure1_schematic}(a)). The specific dimensions of the glass container and elevated platform are also provided in Figure \ref{fig:Figure1_schematic}(a). The photograph of the container, along with the positions of various cameras and illumination panels and its top view, are shown in Figure \ref{fig:Figure1_schematic}(b) and (c), respectively. The illumination was provided by flat light-emitting diode (LED) panels. A solenoid (35V, 0.6A, capable of holding up to 1.5kg) holds the projectile, and its release triggers various cameras. 

\begin{figure}
\centering
\includegraphics[width=0.7\linewidth]{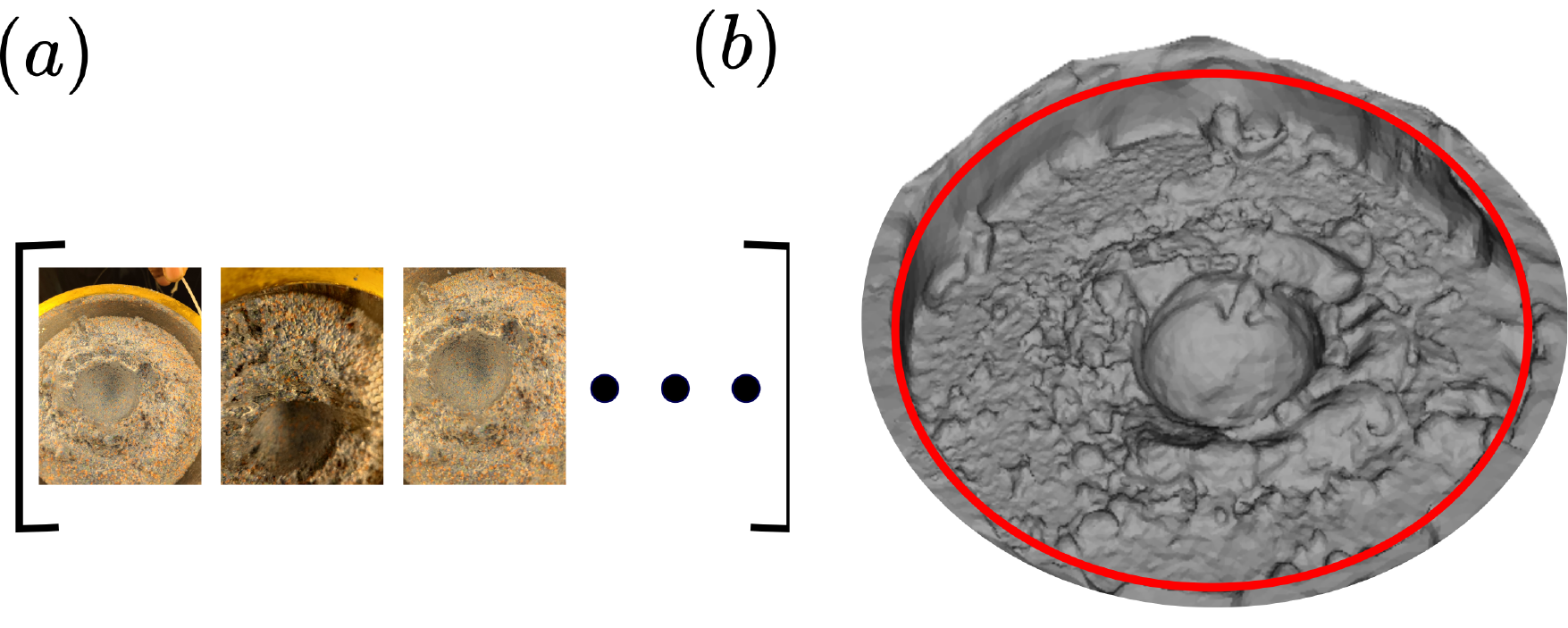}
\caption{Steps in the photogrammetry-based reconstruction of the crater: (a) About 100 images are captured from various angles, and (b) the tessellated surface generated from the point cloud of the coordinates obtained from the Poly-Cam Software \citep{polycam}. The red circle shows the inner wall of the glass cylinder whose diameter is 150 mm.}
\label{fig:Figure_polycam}
\end{figure}

We employ hydrophobic granular material in the form of polyhedral-shaped ``magic sand'' grains obtained from Education Innovation Inc. (USA). These grains are crafted by coating polyhedral-shaped sand particles with a hydrophobic material. Figure \ref{fig:Figure1_schematic}(e) shows the size distribution of the grains with a peak $2r$ at 0.6 mm, where $2r$ is the diameter of a grain. This material had a contact angle of $150^{\circ}$ \citep{karmakargsPhysRevE.95.042903}. Figure \ref{fig:Figure1_schematic}(f) and (g) depict the grains captured at varying levels of magnification through scanning electron microscopy (SEM), revealing the hydrophobic coating and the polyhedral shapes of the grains, respectively.

Water was introduced into the chamber below the platform. This prevents the turbulence generated during the filling process from affecting the surface of the sand. This precaution is essential, as turbulence could otherwise roughen the flat sand surface, making it challenging to determine the crater profile accurately. The surface of the hydrophobic sands attains a distinct lustre when the water covers it. This lustre originates from the total internal reflections of light at the pinned water-air interface \citep{karmakargsPhysRevE.95.042903}.

A float (length 80 mm and diameter 12 mm) is confined to move within a glass tube (diameter 16 mm and height 600 mm). The float is carefully weighed to position its top 1 cm above the water level. A thin piece of metal is attached to the top of the float, which can be monitored using a dedicated camera (Camera 3: Raspberry Pi HQ camera with a 16mm telephoto lens). With this arrangement, we track the movement of the air-water interface, denoted as $\partial \mathcal{A}$. The overall displacement of $\partial \mathcal{A}$, termed as $\Delta (\partial \mathcal{A})$ is then used to calculate the volume of air released after subtracting the volume of the projectile. The details of this method are given in Appendix \ref{AppxA}.

In the experiments, we used smooth stainless steel spherical projectiles (density 7600 kg/m$^3$) with diameter, $D_p =$ 7.5 mm, 15 mm, 12.5 mm, 20 mm, 40 mm, 50 mm and 60 mm. The projectiles become rusted when exposed to air and the resultant roughness of the surface significantly affects the dynamics of the trapped air. To minimise this, we store the projectiles in oil, and a thorough cleaning process with acetone is employed to eliminate any unwanted materials adhering to the surface before and after each experiment.

Figure \ref{fig:Figure1_schematic}(d) shows typical images of the escaping bubbles after impact in the $yz$ and $xz$ planes captured using cameras 1 and 2 (Raspberry Pi Camera V2), respectively. The images were then utilized to analyze the motion of the projectile as it entered the liquid, as well as to assess the shape and size of the bubbles emerging post-impact. The shape of these bubbles, resulting from the impact of the projectiles, is significantly more complex than those of typical air bubbles rising in undisturbed liquids \citep{tripathi2015dynamics,sharaf2017shapes}. The cratering dynamics were recorded using a high-speed camera (Phantom Miro M310) at a consistent interval of $2000$ frames per second (fps), which is positioned (not shown) to directly observe the granular surface $\partial \mathcal{B}$ to capture the impact dynamics. The impact velocity is calculated by post-processing the high-speed images.

The cylinder containing sand was raised using the pulley system after each impact experiment. The size distribution of the sand grains is depicted in Figure \ref{fig:Figure1_schematic}(e). The 20 mm water column above the sand aids in preserving the shape of the crater. After removing the projectile with a magnet, we captured around 100 images of the sand bed in the presence of the 20 mm water column. These images were then analyzed using Poly-Cam, a photogrammetry software \citep{polycam} to generate the crater's surface profile. This approach yields a point cloud representing surface coordinates, which is typically analyzed by fitting it to different functional forms as outlined in Table 1 of \cite{pacheco2017craters}, depending on the mechanism of crater formation. In our case, we have fitted the obtained point clouds to part of a sphere. The steps of this process and its results are shown in Figure \ref{fig:Figure_polycam}(a,b). After completion of the experiment, water was evacuated using a submersible pump positioned beneath the elevated platform, and the bed was re-prepared using the previously described procedure.

\section{Results and discussion} \label{sec:dic}

During its descent, the projectile interacts with two interfaces. The first interaction occurs at the air-water interface $(\partial \mathcal{A})$, and the second occurs when the projectile reaches the granular bed $(\partial \mathcal{B})$. The first interaction causes a deceleration of the projectile. This is a well-studied problem where the competition between viscous and surface tension forces determines the criteria for cavity formation and air entrapment \citep{truscott2014water}. We will briefly touch upon this well-studied topic to establish the relevant parameters for our problem for the sake of completeness. Subsequently, we discuss the crater formation process due to the impact of the projectile with the granular bed ($\partial \mathcal{B}$), which is the primary focus of this study.

\begin{figure}
\centering
\includegraphics[width=0.9\linewidth]{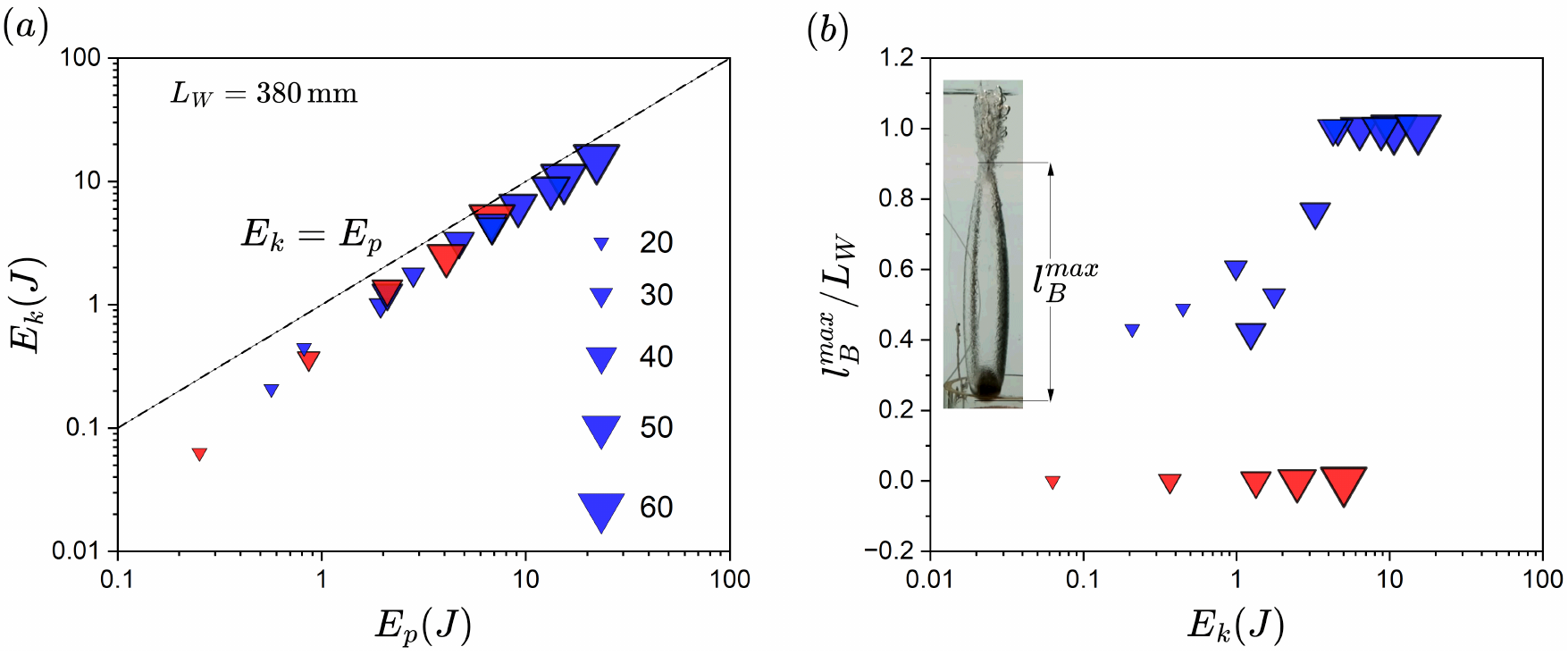}
\caption{(a) Variation of the kinetic energy ($E_k$) with the potential energy ($E_p$) of the projectile. The dash-dotted line represents $E_k=E_p$ for $L_W=380$ mm. (b) Variation of normalised the maximum length of the air cavity ($\ell_B^{max}/L_W)$ as a function of the kinetic energy ($E_k$) of the projectile. The size of the symbol is proportional to the diameter $D_p$ of the projectile (mm) as shown in panel (a). The blue and red symbols correspond to cases $\ell_B^{max} \ne 0$ and $\ell_B^{max} \approx 0$, respectively. The inset in panel (b) shows the image of a cavity behind the projectile showing $\ell_B^{max}$.}
\label{fig:Fig_partial_A}
\end{figure}

Due to the higher viscosity of water than air, the projectile undergoes significant deceleration upon entering the liquid pool. Figure \ref{fig:Fig_partial_A}(a) depicts the variation of kinetic energy ($E_k$) with the potential energy ($E_p$) of the projectile. As the projectile migrates within the liquid pool, its velocity decreases due to the viscous and cavity drag forces \citep{glasheen1996vertical}. Consequently, the kinetic energy measured just before the projectile impacts the bed $(\partial \mathcal{B})$ is consistently smaller than its potential energy. This deceleration effect is more pronounced for low-energy projectiles with smaller values of $D_p$. The entry of projectiles into a liquid pool commonly leads to the formation of air-entraining cavities. The length $\ell_B$ of the cavity increases over time, and when surface energy dominates over viscous effects, it undergoes pinch-off. The maximum extent of the cavity is denoted by $\ell_B^{max}$. The cavity closure events (pinch-offs) can be observed in various scenarios. A surface seal occurs when the cavity closes on the free surface ($\partial \mathcal{A}$). If the projectile contacts the granular bed before the cavity closes ($\ell_B^{max}\simeq O(L_W)$), a surface seal scenario is realized. In contrast, a deep seal is characterized by an initial pinch-off inside the liquid pool without the cavity extending to the top. If the projectile reaches the bed after the cavity has closed, a deep seal event is recorded ($\ell_B^{max}< L_W$). 

Figure \ref{fig:Fig_partial_A}(b) illustrates the change in length of the normalized air cavity ($\ell_B^{max}/L_W$) with the variation in kinetic energy ($E_k$) of the projectile. Here, the blue-coloured symbols are used for cases where air is entrained with the projectile ($\ell_B^{max} > 0$), while red-coloured symbols denote cases where no air is entrained ($\ell_B^{max} \sim 0$). For $E_k > 4$ J, cavity formation and surface sealing events are observed. However, below this energy threshold, two scenarios emerge. In the first case, the air is entrained, and a deep pinch-off occurs ($0<\ell_B^{max}<L_W$). In the second case, no air-entraining cavities are formed ($\ell_B^{max} \sim 0$). This is not surprising, given that the phenomenon of air entrainment is significantly influenced by the surface properties of the projectile \citep{truscott2014water}. 

\begin{figure}
\centering
\includegraphics[width=0.9\linewidth]{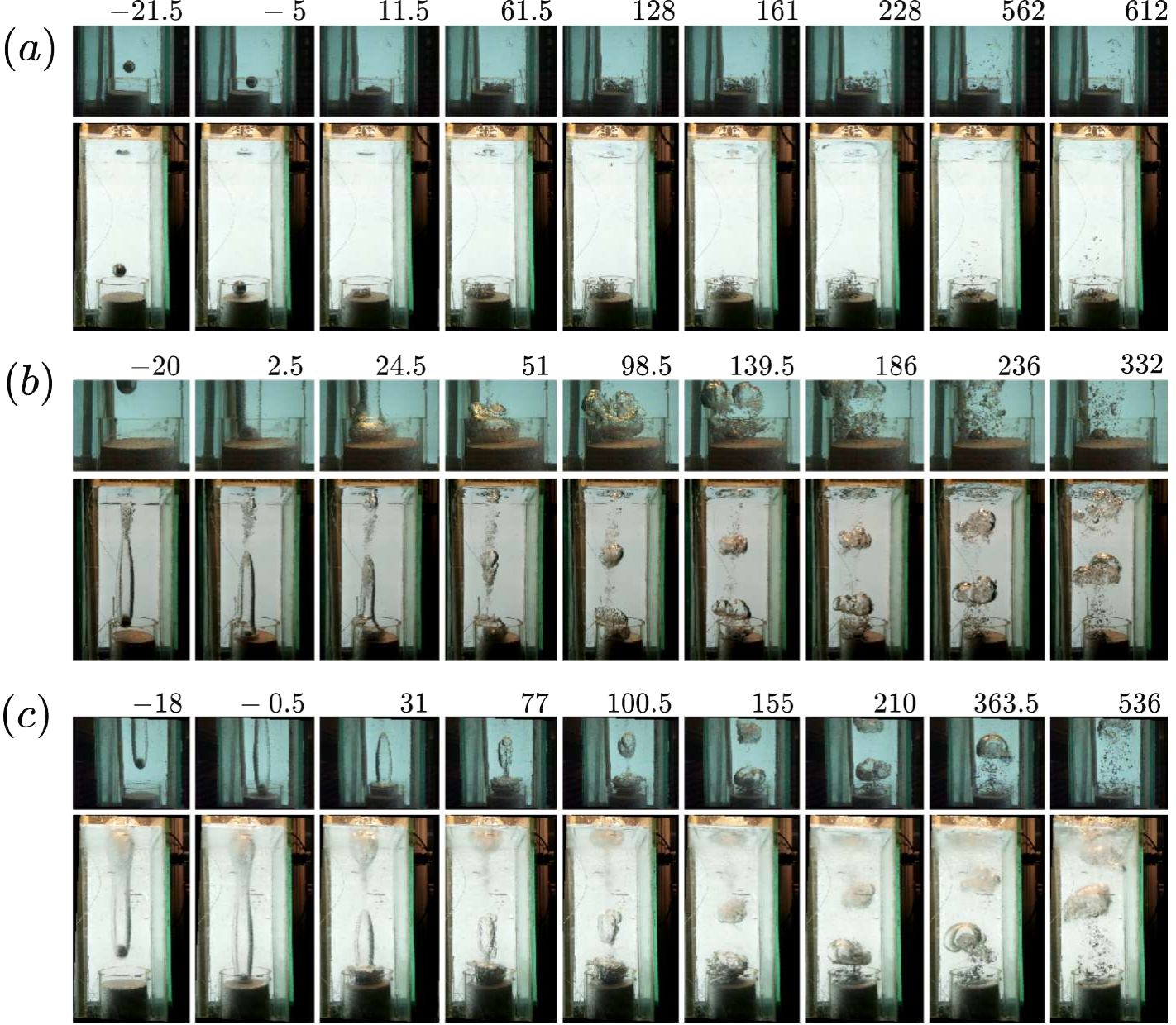}
\caption{Impact dynamics of the projectile ($D_p=40$ mm) on the granular bed submerged in water for $(a)$ $E_k=1.3$ J ($\ell_B^{max}=0$), $(b)$ $E_k=3.3$ J (pinch-off of the air bubble in the bulk liquid; $0 < \ell_B^{max} < L_W$) and $(c)$ $E_k=4.6$ J (surface pinch-off at $\partial \mathcal{A}$; $\ell_B^{max} \approx O(L_W)$). In each panel, the top image depicts the dynamics at the bed, and the bottom shows the dynamics over the entire liquid column captured using different cameras. The time, $t$ in milliseconds (ms), is shown at the top of each panel, with $t = 0$ ms representing the instant when the projectile touches the granular bed surface.}
\label{fig:montage}
\end{figure}

Next, we explore the impact of the projectile on the underwater granular bed $(\partial \mathcal{B})$. Three distinct scenarios arise for such impacts on the hydrophobic granular bed: (i) a projectile without entrained air, (ii) a projectile with entrained air and a deep-seal cavity closure, and (iii) a projectile with entrained air and a surface-seal cavity closure. In Figure \ref{fig:montage}, we sketch these scenarios for three representative values of $E_k$. For the cases where $\ell_B^{max} \simeq 0 $ (see Figure \ref{fig:montage}(a)), no air is entrained with the projectile. After the collision of this projectile with the sand bed ($\partial \mathcal{B}$), we observe a cluster of small spherical-cap bubbles of air escape from the sand bed. For the cases with $\ell_B^{max} > 0$ (refer to Figure \ref{fig:montage}(b) and (c)), we distinguish three stages in the cavity development process. These stages include the elongation of the cavity as the projectile advances within the water, the formation of the doughnut-shaped bubble, and the retraction of the doughnut-shaped bubble after fragmenting into satellite bubbles towards the free surface. The doughnut-shaped bubble, accompanied by recirculating flows, emerges due to the abrupt halt of the projectile while the air in the cavity continues moving downward. This generates inertia-induced shear flow, leading to a Kelvin-Helmholtz-type instability. 

\begin{figure}
\centering
\includegraphics[width=0.9\linewidth]{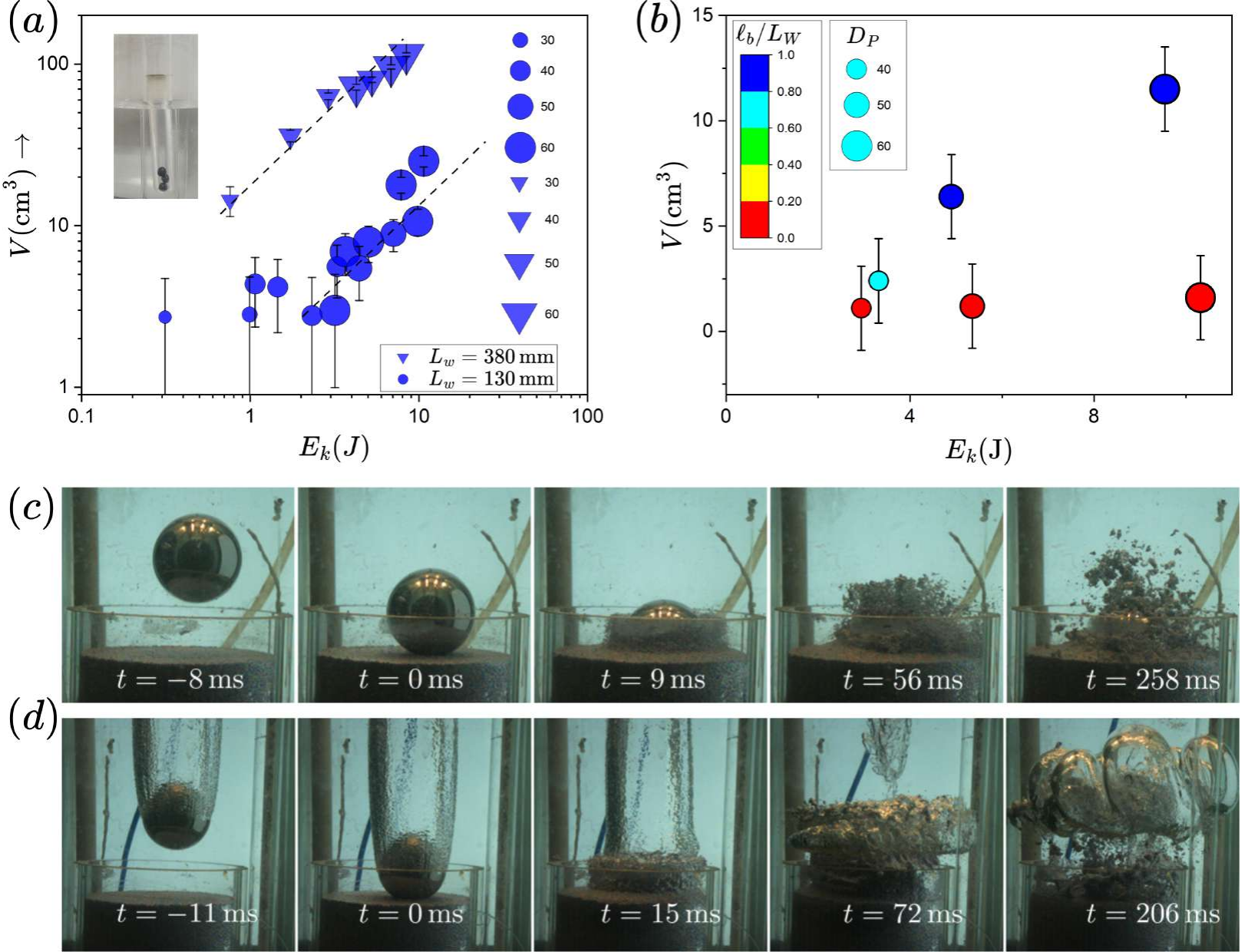}
\vspace{0.2cm}
\caption{(a) Variation of the volume ($V$ in cm$^3$) of air ejected from the submersed granular bed with the kinetic energy, $E_k$ for $L_W = 380$ mm (squares) and $130$ mm (circles). The inset in the figure shows a float employed to monitor the water level, providing a measure of the escaped gas volume. For both the values of $L_W$, $V \sim E_k$ (dotted line). The size of the symbol is proportional to the diameter $D_p$ of the projectile in mm. (b) The volume of gas escaped when $\ell_b^{max}>0$ (with air drag) and $\ell_b^{max}=0$ (without air drag) for $L_W=380$ mm. (c) Snapshots of the impact process for $\ell_b^{max}=0$ and (d) $\ell_b^{max}>0$ using the projectile with $D_p = 60$ mm. Here, $t=0$ refers to the moment of impact.}
\label{fig:Figure_volume}
\end{figure}

The transition between the last two stages occurs after the projectile impacts the bed. As discussed earlier, the impact of the projectile and the resulting shear flows disrupt the arrangement of the granular bed. This temporary disturbance creates gaps in the granular skin, allowing the air trapped within the pore spaces to escape. In Figure \ref{fig:Figure_volume}(a), we plot the air volume released from the cracked granular bed $V$ for different impact energies $E_k$. We find $V$ to be proportional to $E_k$. The hydrostatic pressure due to the water column destabilizes the air trapped in the pore spaces of hydrophobic sand. Thus, for the same impact energy $E_k$, the amount of air released $V$ from the sand bed increases with the water column height $L_W$, as shown in Figure \ref{fig:Figure_volume}(a). 

Rough projectiles induce greater air entrainment compared to smoother ones. We use this feature to vary the maximum cavity length ($\ell_b^{max}$) for projectiles of the same size and energies and investigate the role of shear flows in the volume of gas ($V$) released from the sand bed. While $\ell_b^{max}$ does not measurably affect crater dimensions ($D_C$ and $h$), the quantity of air released from the sand bed depends on both $E_k$ and $\ell_b^{max}$. Figure \ref{fig:Figure_volume}(b) illustrates the volume $V$ of gas released for projectiles of the same $D_P$ and $E_k$, for different values of $\ell_b^{max}$. We observe that for smooth projectiles, both $\ell_b^{max}$ and $V$ are approximately zero. In contrast, for rougher balls, $\ell_b^{max}>0$ and $V$ increases linearly with $E_k$.

Figures \ref{fig:Figure_volume}(c) and (d) illustrate the impact process for smooth and rough projectiles, respectively, where $D_P=60$ mm and $E_k \approx 10$ J. Smooth projectiles have $\ell_B^{max}=0$, while rough projectiles have $\ell_B^{max}>0$. In Figure \ref{fig:Figure_volume}(c), where $\ell_B^{max}=0$, the impact results in the ejection of a small amount of sand particles. These particles trap minimal air, which is subsequently released. Conversely, Figure \ref{fig:Figure_volume}(d) illustrates that when the projectile with the cavity impacts the bed, it ejects a small amount of sand and expands the cavity into a growing doughnut-shaped bubble. The increased volume of air released from the bed in impacts with a finite cavity supports our assertion that the inertia-driven shear flow induced by the doughnut-shaped bubble significantly influences the mechanism of trapped air release. We note that there exists a critical energy ($E_k \sim 0.1$ J) of the projectile below which no bubbles are released from the bed due to impact. This suggests the presence of energy barriers.
 
\begin{figure}
\centering
\includegraphics[width=0.95\linewidth]{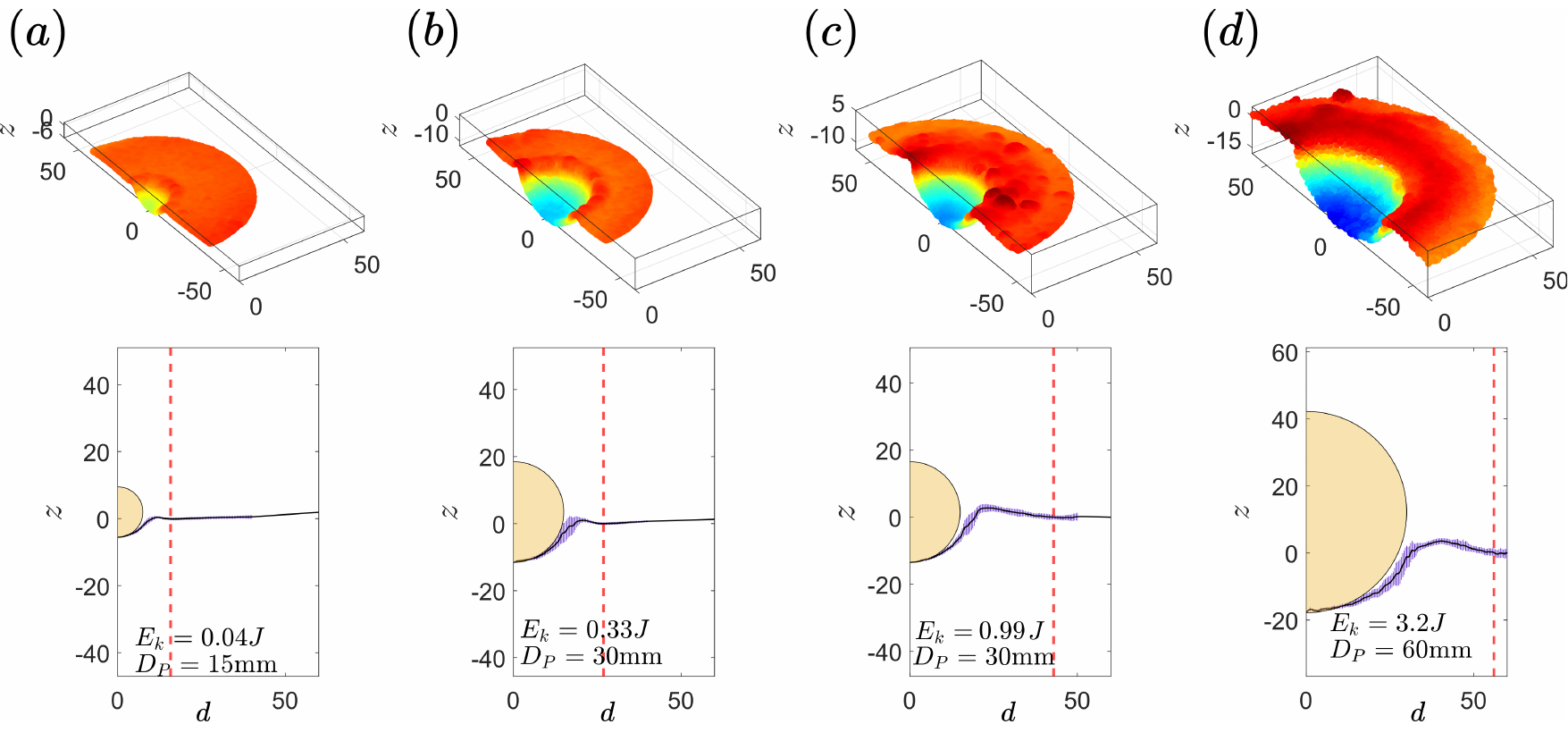}
\caption{The cross-sectional view (top panel) and the azimuthally averaged crater height (bottom panel) for different values of $E_k$. The red dotted line shows the extent of the crater ($D_C/2$). In the bottom panels, the half circles in yellow represent the projectile used. All dimensions are in mm.}
\label{fig:crater-rim}
\end{figure}

Following each impact, we wait for all the bubbles emerging from the bed to reach the top surface $\partial \mathcal{A}$. Once all the air has escaped and the system reaches a steady state, we extract the cylinder containing the sand, as discussed earlier, and quantitatively examine the morphology of the resulting crater. The typical three-dimensional morphology of the crater, obtained by photogrammetry using the software Poly-Cam \citep{polycam} is shown in Figure \ref{fig:Figure_polycam}(b). In contrast to dry granular matter, where a low-energy cavitation process is associated with material ejection, the presence of skin (a pinned air-water interface) in the present study prevents such ejection. Here, the energy of the incoming projectile is used to deform the skin, which in turn increases its surface area (Figure \ref{fig:statement}). The rims of craters in hydrophobic underwater sand tend to be fragile. 

\begin{figure}
\centering
\includegraphics[width=0.95\linewidth]{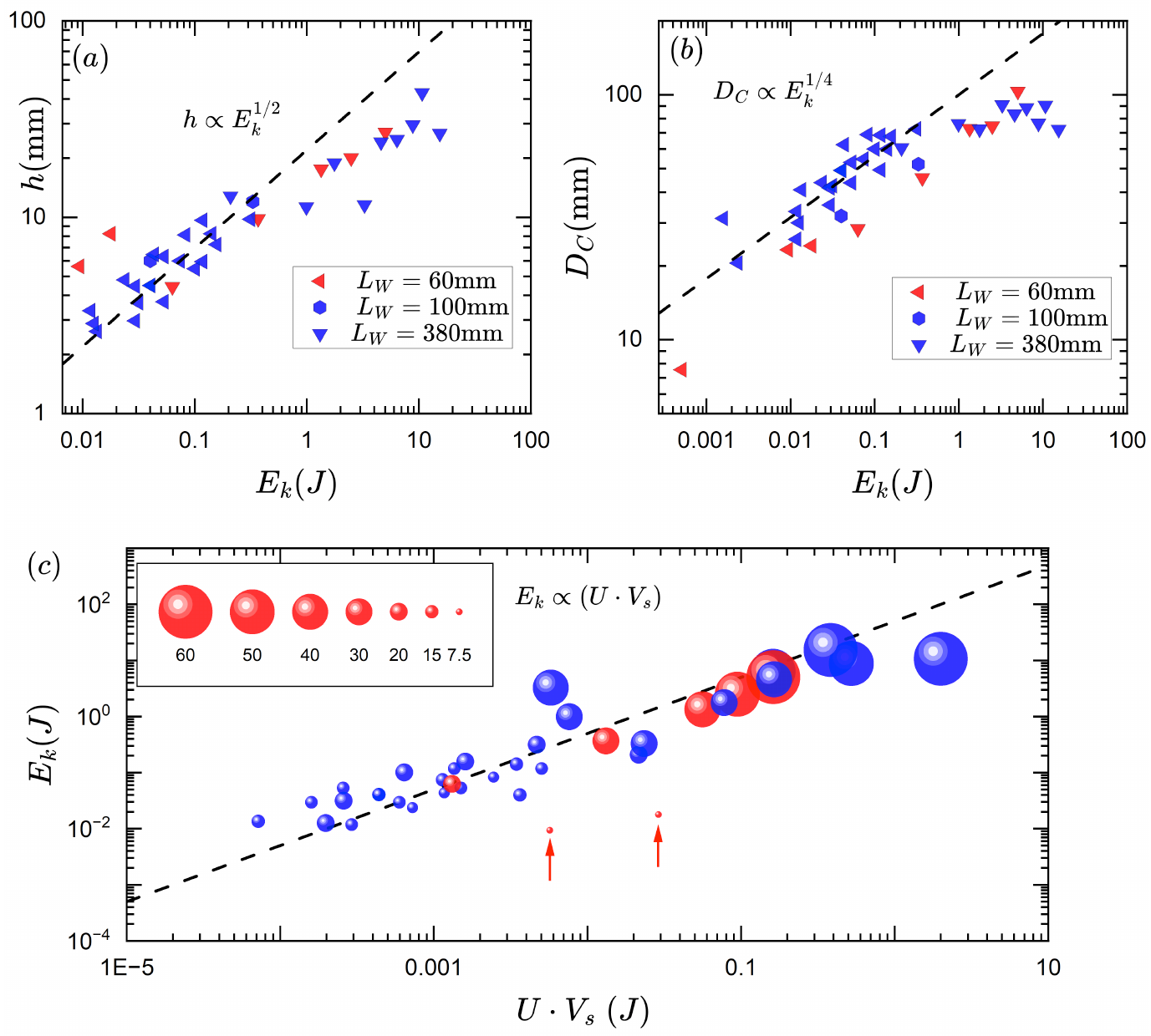}
\caption{Demonstration of scaling laws. Variations of (a) depth $(h)$ and (b) diameter $(D_C)$ with the kinetic energy of the projectile $(E_k)$. Data corresponding to experiments for different values of $L_W$ are represented by distinct symbols. (c) Variation of $E_k$ with $U \cdot V_s$, where $U$ is the strain energy density, and $V_s$ is the volume of the skin over which stress is applied. The dashed line represents a linear relationship between $E_k$ and $U \cdot V_s$. In panel (c), the size of the symbol (shown as the inset in mm) is proportional to the diameter $D_p$ of the projectile. The two red points marked by arrows do not obey this linear relationship. These points correspond to small-diameter projectiles with high values of $E_k$. In all the panels the blue and red symbols correspond to cases $\ell_B^{max} \ne 0$ and $\ell_B^{max} \simeq 0$, respectively.}
\label{fig:figure7_scaleddata}
\end{figure}

In Figure \ref{fig:crater-rim}, the top panels show the three-dimensional (3D) reconstruction of the crater, while the bottom panels show the azimuthally averaged variation of the crater height $z$ as a function of the distance $(d)$ from its centre for different values of $E_k$. The centre is defined to be the deepest point of the crater. This variation shows a peak before becoming flat. We set this flat surface to be $z=0$. The peak corresponds to the crest of the crater, and the distance at which the variation flattens is defined to be $D_C/2$. The difference in height between the deepest point of the crater and the flat surface is defined as $h$. The projectiles used are represented as yellow half-circles. At the point of impact, the lower part of the crater closely conforms to the projectile's surface, leading us to model the crater's surface as part of a sphere, as illustrated in Figure \ref{fig:crater-rim} (bottom panels). This spherical approximation is employed in subsequent analyses of crater formation scaling laws. The assumption that the projectile's shape influences the crater's shape has been previously used in the context of crater formation in cohesive granular media \citep{katsuragi2017physics}. A careful inspection of the rim structure reveals that it exhibits height modulation. Notably, similar undulations have been previously observed in the outer edge of structures formed through the interaction of vortex rings with granular beds. In these cases, the splitting of vortices into substructures has been attributed to the formation of these intricate radial patterns in the granular bed \citep{yoshida2012collision,munro2009sediment}.

In Figure \ref{fig:figure7_scaleddata}(a) and (b), we plot the variation of the crater depth $(h)$, crater diameter $(D_C)$, with the kinetic energy $(E_k)$. Unlike the clearly defined power-law dependencies observed in dry granular media \citep{PhysRevE.68.060301,PhysRevE.90.032208}, these parameters do not exhibit any such dependence on $E_k$. In the process of forming the crater, an initial patch of the skin with an area $A$ extends to $A+\Delta A$. Assuming the crater to have a bowl-shaped spherical depression (a spherical cap), we obtain $A=\pi{D_C}^2/4$. The curved area of the crater is $\pi (h^2 +{D_C}^2/4$) and thus, $\Delta A= \pi h^2$. In this geometry $(D_C/2)^2=D_P h-h^2$. Thus for small $h$, ${D_C}^2 \propto h$. For $E_k<1$ J, we find that $h \propto E_k^{1/2}$ and $D_C\propto E_k^{1/4}$. These scaling laws would imply that a fraction of the energy of the impact went into deforming the surface of the crater, i.e. $\Delta A = \pi h^2 \propto E_k$. Although these relationships are consistent with the spherical geometry of the craters, we can see that both of these do not hold for $E_k>1$ J, as seen in Figure \ref{fig:figure7_scaleddata}(a) and (b). 

To rationalize our observation of the scaling behaviour, we assume that a fraction of the projectile's energy ($E_k$) is used as the strain energy ($U \cdot V_s$) for the skin of the crater. Here, $U=\frac{Y}{2} \left(\frac{\Delta A}{A}\right)^2 =\frac{Y}{2} \left(\frac{2h}{D_C}\right)^4 $ represents the energy density associated with the increase in the area of the skin ($\partial \mathcal{B}$), and and $V_s = A \cdot r$ is the volume of the skin. Here, $Y$ is the effective elastic constant of the skin. Figure \ref{fig:figure7_scaleddata}(c) depicts the variation of the energy of impact $E_k$ as a function of $U \cdot V_s$, under our spherical-cap assumption. The effective elastic constant of the granular skin, $Y$ was independently measured  to be of the order of $10^5$ Pa \citep{karmakargsPhysRevE.95.042903}. The dotted line represents a linear relationship between $E_k$ and $U \cdot V_s$. This observation indicates that about 5\% of the kinetic energy of the projectile is used in deforming the skin $(\partial \mathcal {B})$. These scaling laws hold for projectiles with different diameters and various values of the water height $L_W$ and are independent of the air entrainment scenario. Finally, we point out a caveat that small projectiles with high values of $E_k$ tend to burrow into the bed and form cylindrical cavities. These points, marked by arrows in Figure \ref{fig:figure7_scaleddata}(c), appear as outliers as the spherical cavity assumption fails.

\begin{figure}
\centering
\includegraphics[width=0.95\linewidth]{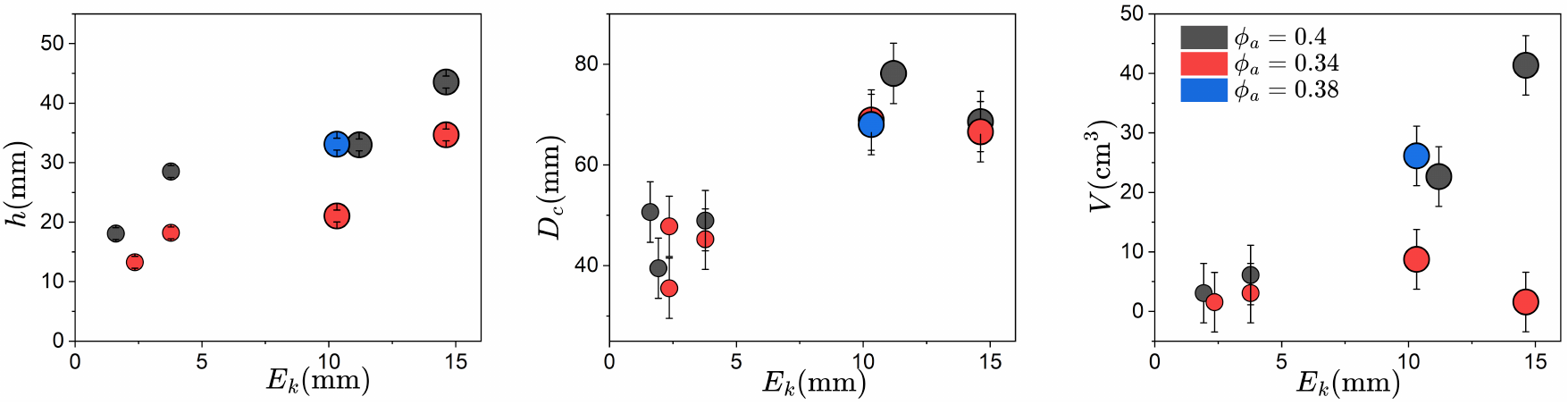}
\caption{Demonstration of the impact of compaction on the sand. The black symbols represent the sand bed poured freely, while red symbols indicate the sand bed prepared with strong compaction. The standard protocol (used throughout the manuscript) is depicted using blue symbols. The variations of height $(h)$, diameter $(D_c)$ and volume of the escaping gas $V$ with the impact energy $(E_k)$.}
\label{fig:figure_review1}
\end{figure}

The sand pouring protocol controls the amount of air present on the top layers. To investigate this effect, we define three protocols. (i) No compaction: In this approach, sand is freely poured into the container, and the container is lightly shaken to create a smooth top surface. The fraction of air in the sand ($\phi_a$) is about 40\% by volume. (ii) Standard compaction: Here, the grains are poured into the container and gently tapped. This method is similar to the coffee tamping process. This method ensures the absence of loose grain clusters within the sand bed and approximately traps 38\% air by volume in the pores. For the data presented in the main paper, this method of compaction was used unless mentioned otherwise. (iii) High Compaction: In this process, standard compact sand is further compressed by repeated pounding. This action lowers the free surface of the sand by 5 mm. The compaction, however, primarily occurs near the surface region. The fraction of air trapped is 34\% by volume in the pores. We keep the water level fixed at $L_W=38~ {\rm cm}$ and the sand at $L_S=6.5~ {\rm cm}$. We find that the non-compacted sand releases more air than the compacted sand across various energy levels. While the radius of the crater changes very little, the depth is consistently more in the case of loose sand than in the case of highly compacted sand. Our standard protocol of tapping the surface lightly till a flat plane is obtained is similar to the results of loosely packed sand. These results are shown in Figure \ref{fig:figure_review1}(a,b,c). In addition, we conducted experiments where we manipulated the compaction level, height, and air cavity size of the sand bed while keeping the impact energy constant. Our findings demonstrate several key points: Firstly when the sand bed exceeds the size of the projectiles, there are no discernible alterations in the outcomes. Secondly, variations in the roughness of the projectile directly affect $\ell_B^{max}$. Specifically, projectiles with a larger $\ell_B^{max}$ release more air for the same energy input. For more details, see Appendix \ref{AppxB}.

\section{Concluding remarks} \label{sec:Conc}

We examine the violent dynamics of a projectile impacting a submerged bed and creating a crater as it penetrates through the water column. The projectile's energy forms the cavity, disturbs particles in the bed, and produces complex flow patterns. Despite this complexity, the system shows well-defined scaling laws. These scaling laws differ from those observed in dry and wet sand beds. The observed scaling laws, we believe, will aid in understanding the release of harmful gaseous bubbles from the bottom of seabeds as a result of impacts of objects by providing fundamental insight into the complex physics of this critical problem. We have used only stainless steel spherical projectiles and hydrophobic sands of one kind. It is indeed possible that the material and geometrical properties of sand particles will influence the capability of the bed to store air. Further, one expects that projectile geometries to influence formation. Thus, one might wonder about the universality of the scaling laws presented in this study. However, the simple nature of the scaling laws, their robustness to various experimental conditions and the fact that these laws can be obtained from a simple model based on interfacial energy provide credence to our investigation, albeit for a limited parameter regime.

\appendix

\section{Calibrating the volume change using a float} \label{AppxA}
A float (length 80 mm and diameter 12 mm) is confined to move within a glass tube (diameter 16 mm and height 600 mm). The float is carefully weighed to position its top 1 cm above the water level. A thin piece of metal is attached to the top of the float, which can be monitored using a dedicated camera (Camera 3: Raspberry Pi HQ camera with a 16 mm telephoto lens). With this arrangement, we track the movement of the air-water interface, denoted as $\partial \mathcal{A}$. The overall displacement of $\partial A$, termed as $\Delta( \partial \mathcal{A})$, is then used to calculate the volume of air released after subtracting the volume of the projectile. To track the position of the thin piece of metal attached to the top of the float, we cropped a suitable portion of the image $I(y,z)$ near the top. This cropped image (of size 10 mm in height and 6 mm in width) was converted into a grey-scale image. An average intensity profile $\langle I(z) \rangle_y$ as a function of $z$ values was constructed from the image. The angular brackets denote averaging the intensity values in the $y$ direction for a fixed value of $z$. The step-like feature in this intensity profile, $\langle I(z) \rangle_y$, denotes the position of the thin piece of metal attached to the top of the float. When an object is immersed in the fluid, the floater moves upward by $\Delta z_{f}$. The red and blue lines in Figure \ref{fig:floater}(a) correspond to the position of the thin piece of metal before and after immersion of an object. The float used for monitoring the water level is depicted in Figure \ref{fig:floater}(b). In Figure \ref{fig:floater}(c), we show the calibration curve illustrating the correlation between the change in the height of the meniscus ($\Delta z_{f}$) and the known volume of the immersed object ($V_k$).

\setcounter{figure}{0} \renewcommand{\thefigure}{A\arabic{figure}}
\begin{figure}
\centering
\includegraphics[width=0.95\linewidth]{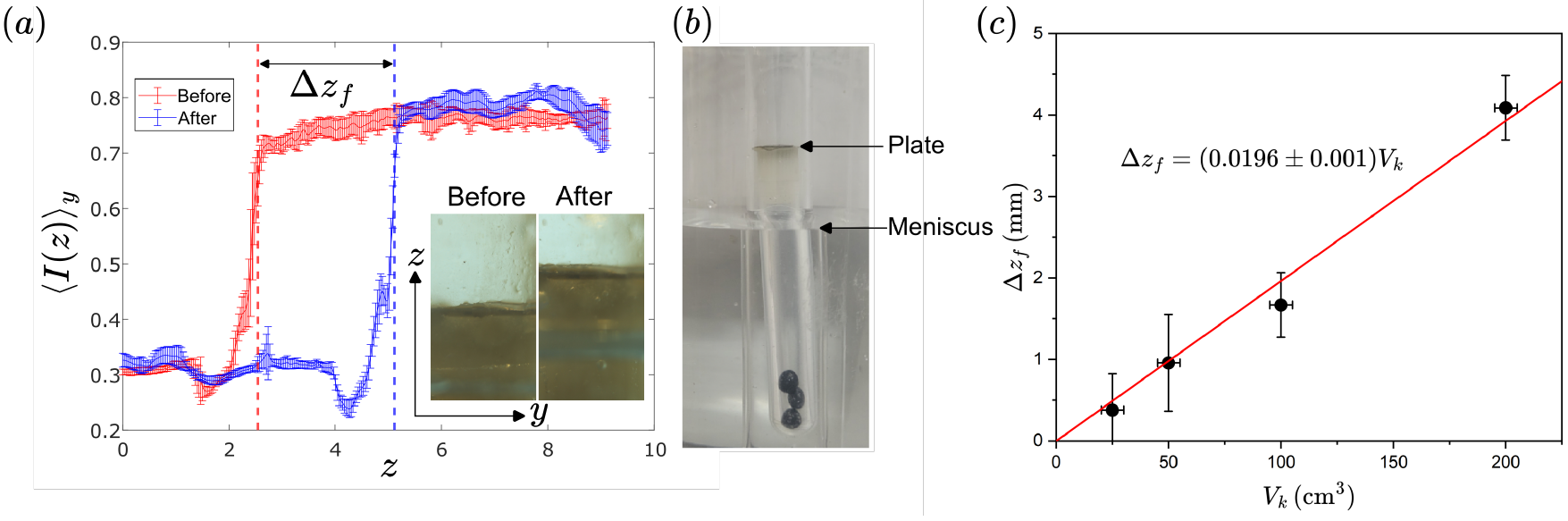}
\caption{(a) Demonstration of the movement of the float, which tracks the water-air meniscus $\partial \mathcal{A}$ before and after a typical impact (see insets). The position of the top of the float is constructed by plotting the average intensity profile $\langle I(z) \rangle_y$ as a function of $z$. Here, the red and blue lines correspond to the images taken before and after the impact. (b) The float used for monitoring the water level $\partial \mathcal{A}$. (c) Calibration curve illustrating the correlation between the change in the height of the meniscus ($\Delta z _{f}$) and the known volume of an immersed object ($V_k$).}
\label{fig:floater}
\end{figure}

\section{Effect of sand height} \label{AppxB}
The impact of the projectile induces a dynamic compaction of the granular bed, with contact line forces opposing this compaction. To test the effect of the height of the sand column on the geometric properties of the crater and the volume of gas released, we perform experiments for different impact energies $E$ for three different values of sand height $L_S$. These results are summarized in Table \ref{tab:kd}. For the range of sand column heights covered in the experiments, the effect of $L_s$ on various parameters is minimal. It is to be noted that for all these cases, $L_s$ was greater than the radius of the projectile $D_p/2$. We have not done experiments in the limit where $L_s <D_p/2$, for in such cases, the mechanical integrity of the sand bed was compromised.

\begin{table}
\centering
\begin{tabular}{cccccc} 
$L_S$(mm) & $D_P$(mm) & $E(J)$ & $h$(mm) & $D_C$(mm) & $V$(cm$^3$)\\ 
130 & 40 & 	2.3 & 	18 & 	37 & 	3 \\ 
65 & 40 & 	2.3 & 22 & 	34 & 	3.5\\ 
45	 & 40 & 	 2.1	 & 12 & 	38 & 	3\\ 
130 & 60 & 		8.2 & 	26 & 	50 & 	43\\ 
65 & 60 & 	8.2	 & 25 & 	49 & 	44\\ 
45	 &60 & 		7.6	 & 25 & 	55 & 	38\\ 
\end{tabular}
\caption{Variation of the geometric parameters, namely the diameter $(D_c)$ and height $(h)$ of the crater and the volume $V$ of the gas released as a function of the impact energy $E$ for three different values of sand height $L_S$. Here, results from projectiles of two diameters $D_p$ have been reported.}
\label{tab:kd}
\end{table}

\hspace{0.5cm}

\noindent{\bf Acknowledgement:} We acknowledge support of the Department of Atomic Energy and Science \& Engineering Research Board, Government of India, under Projects 12-R\&D-TFR-5.10-0100 and CRG/2020/000507. We thank the anonymous reviewers for their valuable feedback. \\

\noindent{\bf Declaration of interests:} The authors report no conflict of interest.



\end{document}